\documentclass[cits]{PoS}
\usepackage{graphicx}
\usepackage{amsmath}
\usepackage{txfonts}
\usepackage{amssymb}
\usepackage{subfig}

\title{Morphologies of low-redshift AGN host galaxies: what role does AGN luminosity play?}

\ShortTitle{Morphologies of low-redshift AGN host galaxies}

\author{\speaker{Carolin Villforth}\thanks{}\\
        Department of Astronomy, University of Florida, Gainesville, FL 32611-2055, USA\\
        E-mail: \email{villforth@astro.ufl.edu}}
\author{Fred Hamann\\
       	Department of Astronomy, University of Florida, Gainesville, FL 32611-2055, USA}
\author{Anton Koekemoer\\
       	Space Telescope Science Institute, 3700 San Martin Dr., Baltimore, MD, 21218, USA}
\author{David Rosario\\
       	Max-Planck-Institut f\"{u}r Extraterrestrische Physik (MPE), Garching, Germany}
\author{Timothy Hamilton\\
       	Department of Natural Sciences, Shawnee State University, Portsmouth, USA}
\author{Elizabeth J. McGrath\\
       	Department of Physics and Astronomy, Colby College, Waterville, ME 04901, USA}
\author{Arjen van der Wel \& YuYen Chang\\
		 	Max Planck Institute for Astronomy, K\"{o}nigstuhl 17, 69117 Heidelberg, Germany}
\author{Yicheng Guo\\
			University of Massachusetts, Department of Astronomy, 710 North Pleasant Street, Amherst, MA, USA}
\author{The CANDELS Collaboration}

\abstract{Mergers of galaxies have been suspected to be a major trigger of AGN activity for many years. However, when compared to carefully matched control samples, AGN host galaxies often show no enhanced signs of interaction. A common explanation for this lack of observed association between AGN and mergers has often been that while mergers are of importance for triggering AGN, they only dominate at the very high luminosity end of the AGN population. In this study, we compare the morphologies of AGN hosts to a carefully matched control sample and particularly study the role of AGN luminosity. We find no enhanced merger rates in AGN hosts and also find no trend for stronger signs of disturbance at higher AGN luminosities. While this study does not cover very high luminosity AGN, we can exclude a strong connection between AGN and mergers over a wide range of AGN luminosities and therefore for a large part of the AGN population.}

\FullConference{The Extreme sky: Sampling the Universe above 10 keV - extremesky2009,\\
		November 6-8, 2012\\
		Max-Planck-Insitut f\"ur Radioastronomie (MPIfR), Bonn,  			Germany}

\begin{document}

\section{Introduction}

A strong connection between AGN and mergers of galaxies has been discussed for a long time, see e.g. \cite{sanders,can_stockton} and has been popular with both observers and theorists, e.g. \cite{hopkins} and references therein. Observations of lower-redshift AGN host galaxies seem to imply that they are often connected to violent early phases of mergers (e.g. \cite{sanders,can_stockton,veilleux}), and this has also been confirmed for certain subsamples of the AGN population (e.g. \cite{urrutia}). Additionally, the finding that the masses of super-massive black holes in local galaxies correlate surprisingly well with other properties of their hosts (e.g. \cite{gebhardt,graham,haering}) has strengthened the idea that AGN and their host galaxies are likely tightly connected. In the last years, it has also become clear that so-called AGN feedback can be used to alleviate certain problems in models of galaxy evolution by quenching starformation in massive galaxies (e.g. \cite{rachel_sam}). 

However, with the availability of deep field data that provide not only high resolution images of AGN hosts but also ample availability of control samples, several studies have attempted to quantify the merger rates in AGN host galaxies compared to control galaxies of the same mass. It has been found found that AGN hosts show no more signs of mergers than the general galaxy population and appear relatively quiescent \cite{cisternas_1,cisternas_2,kocevski,boehm}. This can be explained as being caused by the fact that the majority of AGN studied are often of rather low luminosity, where merger triggering might not be dominant \cite{hopkins_five}. Much of this work has also been performed using human classifiers to identify mergers and disturbed galaxies, a method that might miss more subtle differences in morphological disturbances between populations. Here, we present morphological analysis of low red-shift (z=0.5-0.8) low to moderate luminosity AGN in Chandra Deep Field South using WFC3 $H/F160W$ CANDELS \cite{candels_anton,candels_norm} imaging data to asses the differences between the morphologies of AGN host galaxies and control, in particular, we will compare morphological disturbances as a function of AGN luminosity to see if higher luminosity AGN are more strongly connected to mergers than their lower luminosity counterparts.

\section{Sample, Data \& Analysis}

\begin{figure}
\begin{center}
\includegraphics[width=7.5cm]{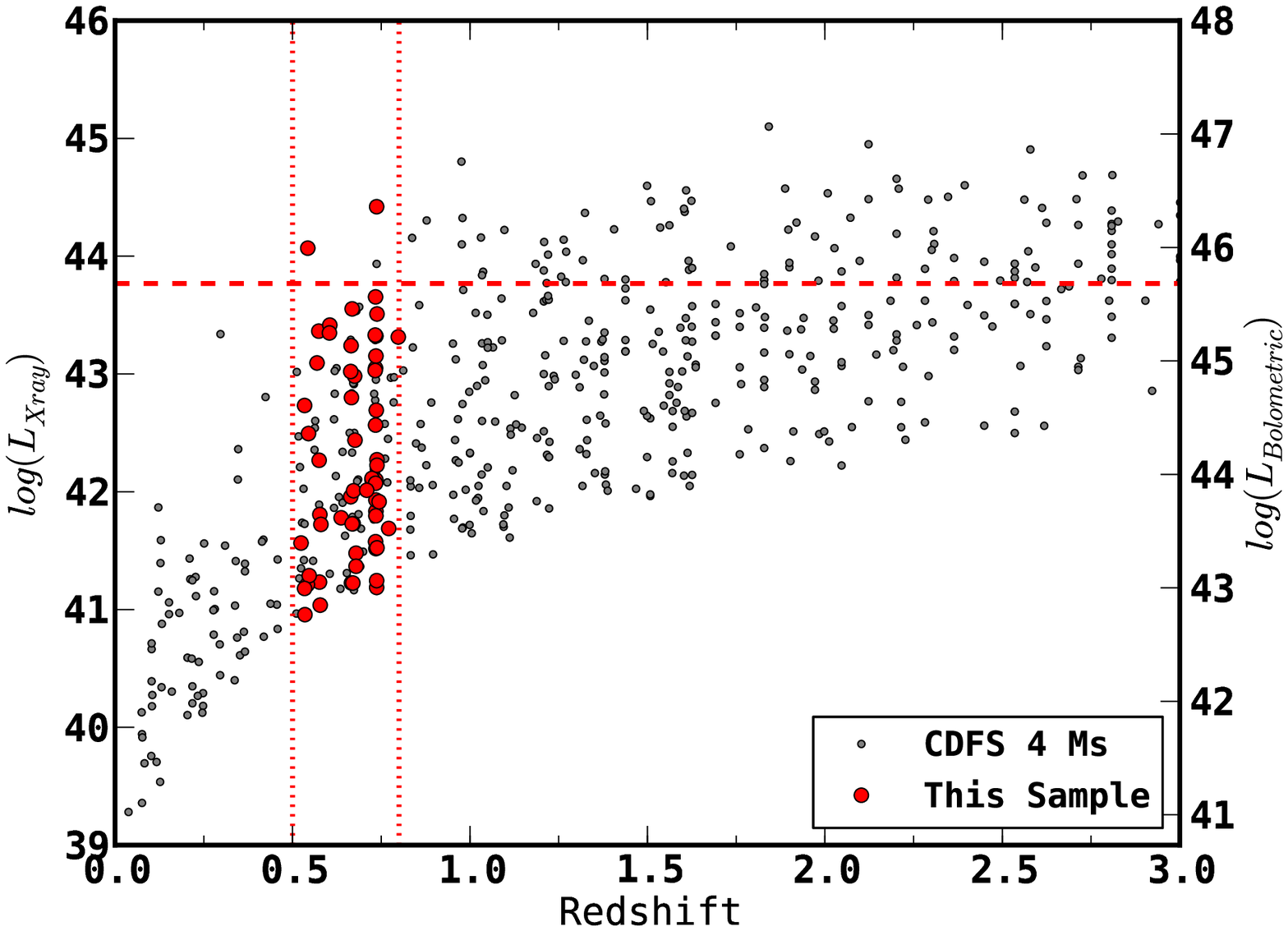}
\includegraphics[width=7.5cm]{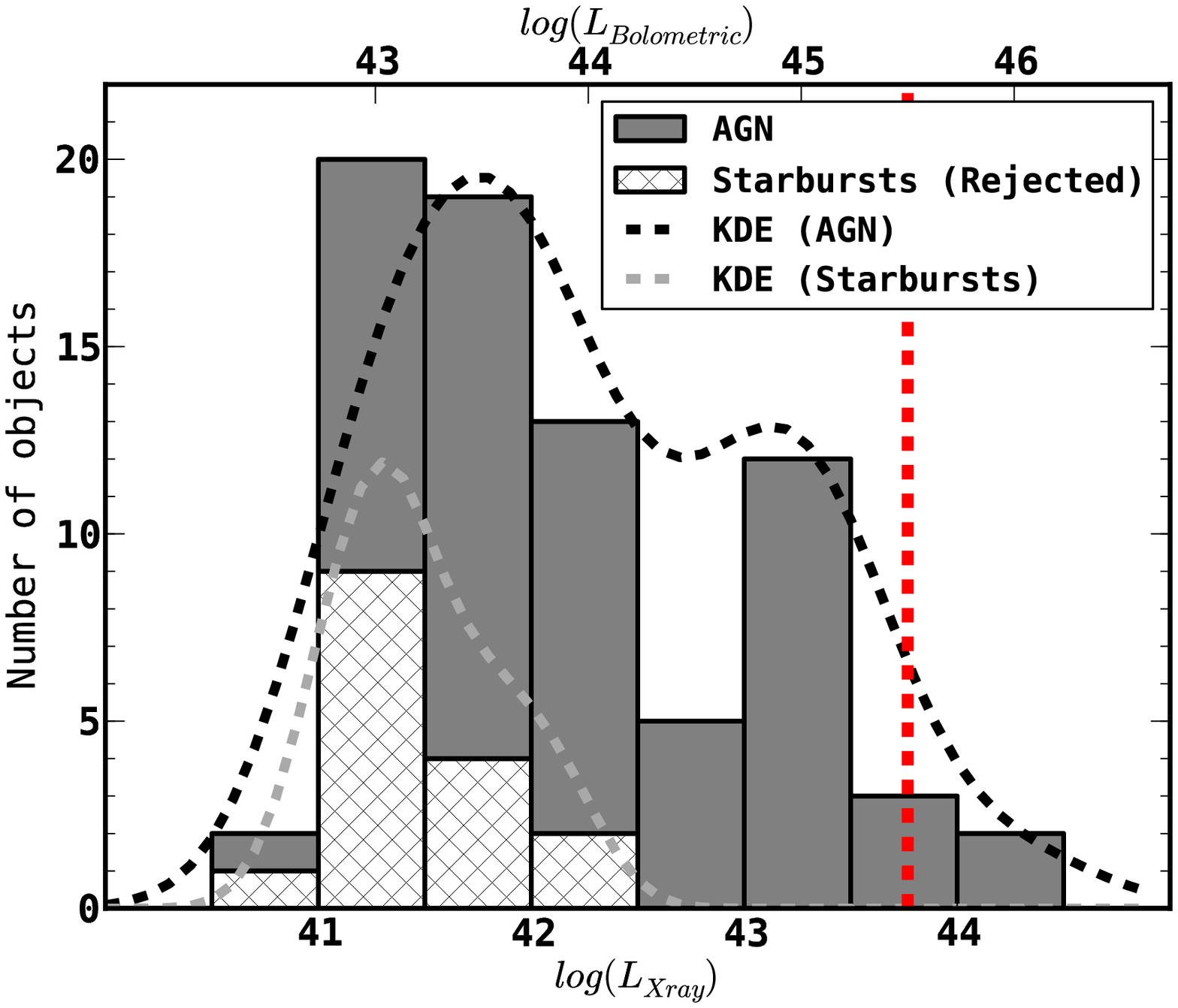}
\caption{Sample of Xray-selected AGN. Left: Xray luminosities and redshifts of all sources in CDFs from Xue et al. 2011. The objects covered in this study are shown as red filled circles. The left axes show the absorption corrected 2.-8kev Xray luminosities from the Xue CDFS 4Ms catalogue, the right axes shows bolometric luminosities calculated using a single bolometric correction for the entire sample. Right: Histogram and kernel density estimator of all xray-sources in our redshift range as well as rejected starbursts.}
\label{F:sample}
\end{center}
\end{figure}

The sample is selected from the Chandra Deep Field South (CDFS) 4 Ms data \cite{xue_chandra_2011}, see Fig. \ref{F:sample}. A redshift range of z=0.5-0.8 is chosen to cover a maximum amount of dynamical range in Xray luminosity while not covering a large enough redshift range to have significant cosmological evolution within the sample and keeping surface brightness dimming minimal, see Fig. \ref{F:sample}). Throughout the paper, we use absorption corrected rest-frame 0.5-8keV luminosities in erg/s from \cite{xue_chandra_2011}. From the 4Ms CDFS Sample, we study all objects covered in CANDELS \cite{candels_anton,candels_norm}. Objects with upper limits only are not included in the sample, this leaves 76 objects with detections covered by CANDELS. Additionally, starbursts are rejected from the sample. This leaves a sample of 66 AGN in the field. The histograms of the AGN as well as rejected starbursts are shown in Fig. \ref{F:sample}.

\begin{figure}
\begin{center}
\includegraphics[width=7.5cm]{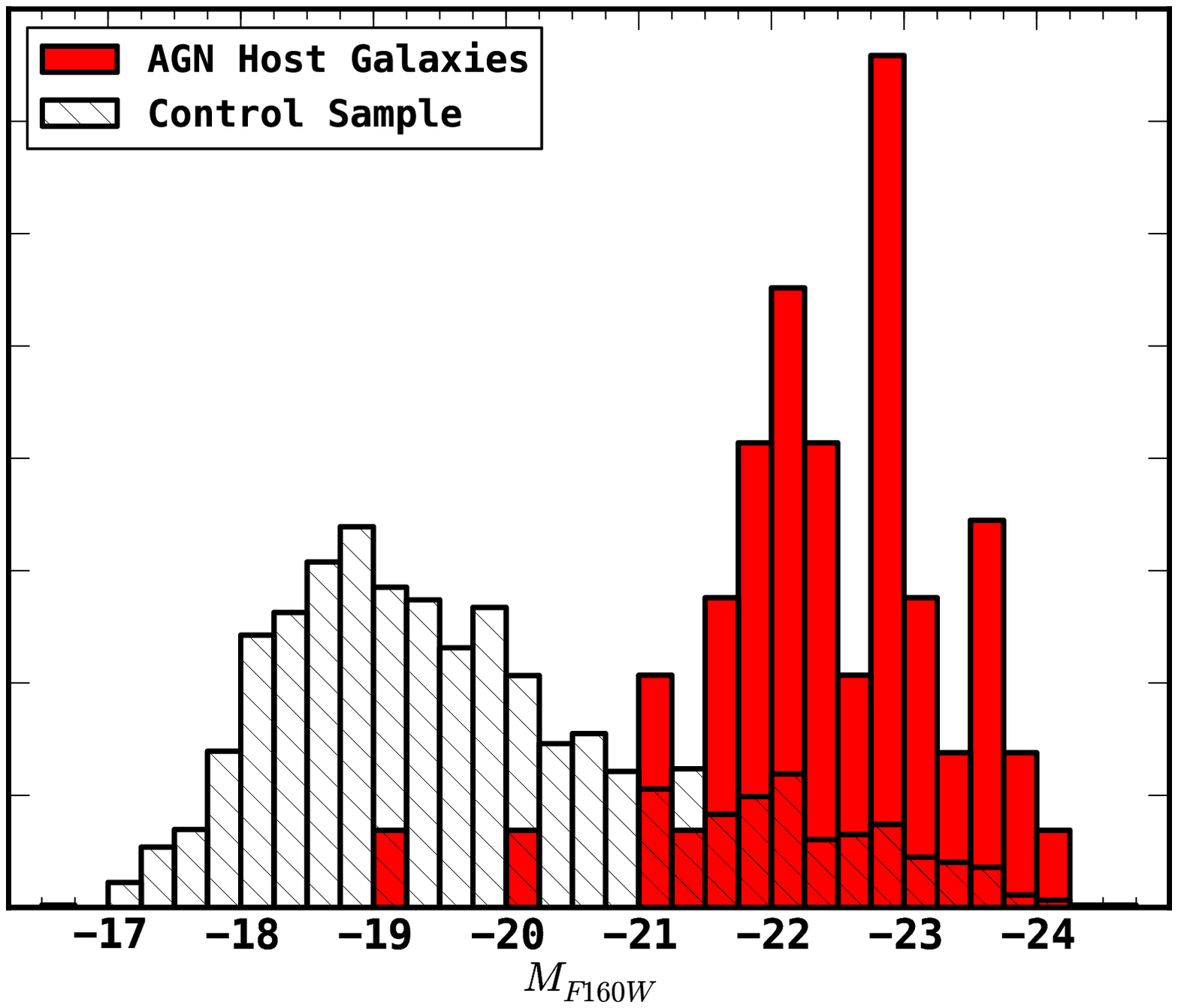}
\includegraphics[width=7.5cm]{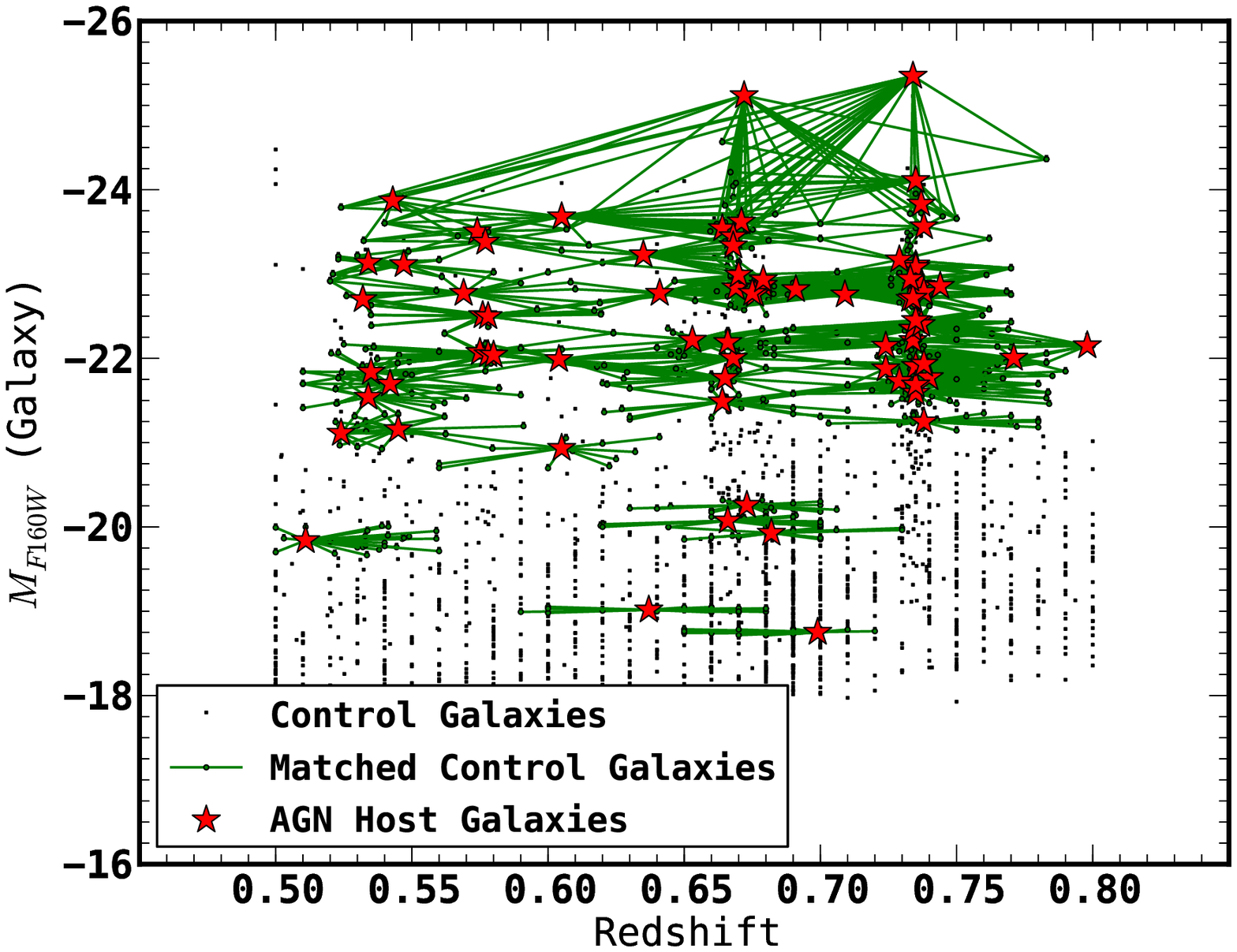}
\caption{Control sample: left panel shows the normed histogram of the AGN host galaxies (red) and all control galaxies in the same redshift range (hashed). Right: absolute $H$ band magnitudes of X-ray selected AGN (red stars) as well as matched control galaxies. Green lines show control galaxies matched to a certain AGN host galaxy.}
\label{F:galaxies}
\end{center}
\end{figure}

The "full" control galaxy sample is chosen to be all galaxies in the sample redshift range covered by CANDELS, AGN are naturally rejected from the sample, but Xray detected starburst galaxies are included in the control sample to avoid biasing the sample against starforming galaxies. For each AGN, we match between 5 and 25 control galaxies as closely as possibly in both redshift and absolute H band magnitude, see Fig. \ref{F:galaxies}. Ideally, matching would be performed in stellar mass (which we will do in an extended version of this study) but $F160W/H$ band magnitude traces stellar mass very well in the redshift range studies, so this is not a problem for our analysis.

Both galaxies and AGN profiles are fit using \textsc{GALFIT} \cite{peng_detailed_2002}. A hybrid PSF combined from TinyTim \cite{tinytim} and empirical derivations of the PSF wings is used. For the AGN  host galaxies, we use a mixture of point source and sersic, if necessary, a second sersic component is added. For the matched control galaxies, we create 'fake' AGN in which a point source with the same magnitude as the matched AGN is added to the galaxy and then subsequently fit using \textsc{GALFIT}. The point source subtracted images of both AGN and fake matched AGN are then used to measure morphology in the AGN host galaxies. To study signs of disturbance, we use the asymmetry A, defined as:

\begin{equation}
A \equiv \sqrt{ \dfrac{ \sum \frac{1}{2} \times (I_{0} - I_{180})^{2} }{ I_{0} ^{2}}  }
\end{equation}

where $I_{0}$ is the image and $I_{180}$ is the image rotated by 180$^{\deg}$ \cite{conselice_2000}. For the purpose of this study, we use segmentation maps to avoid including noise from the background into the measurement. Centering is performed following \cite{conselice_2000}. We have ensured that the algorithm generally reaches a well-defined minimum, visual inspection is performed in addition. 

Due to the point source contamination, some central pixels of the galaxy usually show bad values, while this is also the case for the fake AGN host galaxies, we still do not wish these pixels to dominate the overall asymmetry. Therefore, the central area of all object is masked using a circular mask. We use the same mask for all objects with no dependence on the point source magnitude. This masks also the central areas of galaxies that do not show disturbance by the point source subtractions. However, due to the fact that asymmetry might be different in the central regions and outskirts of galaxies, we do not wish to have different mask regions at different point source magnitudes since this might bias the results.

\section{Results}

\begin{figure}
\begin{center}
\includegraphics[width=7.5cm]{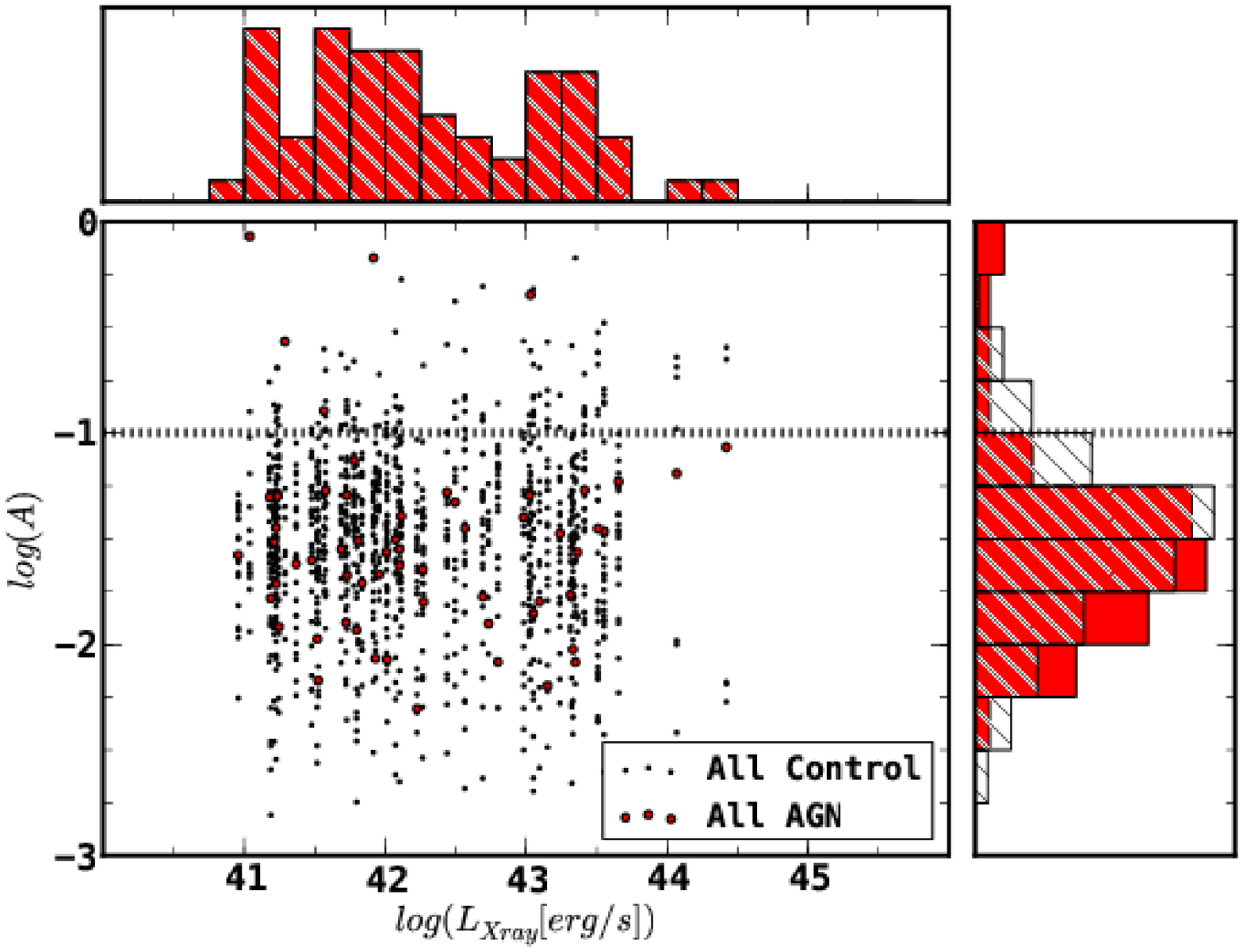}
\includegraphics[width=7.5cm]{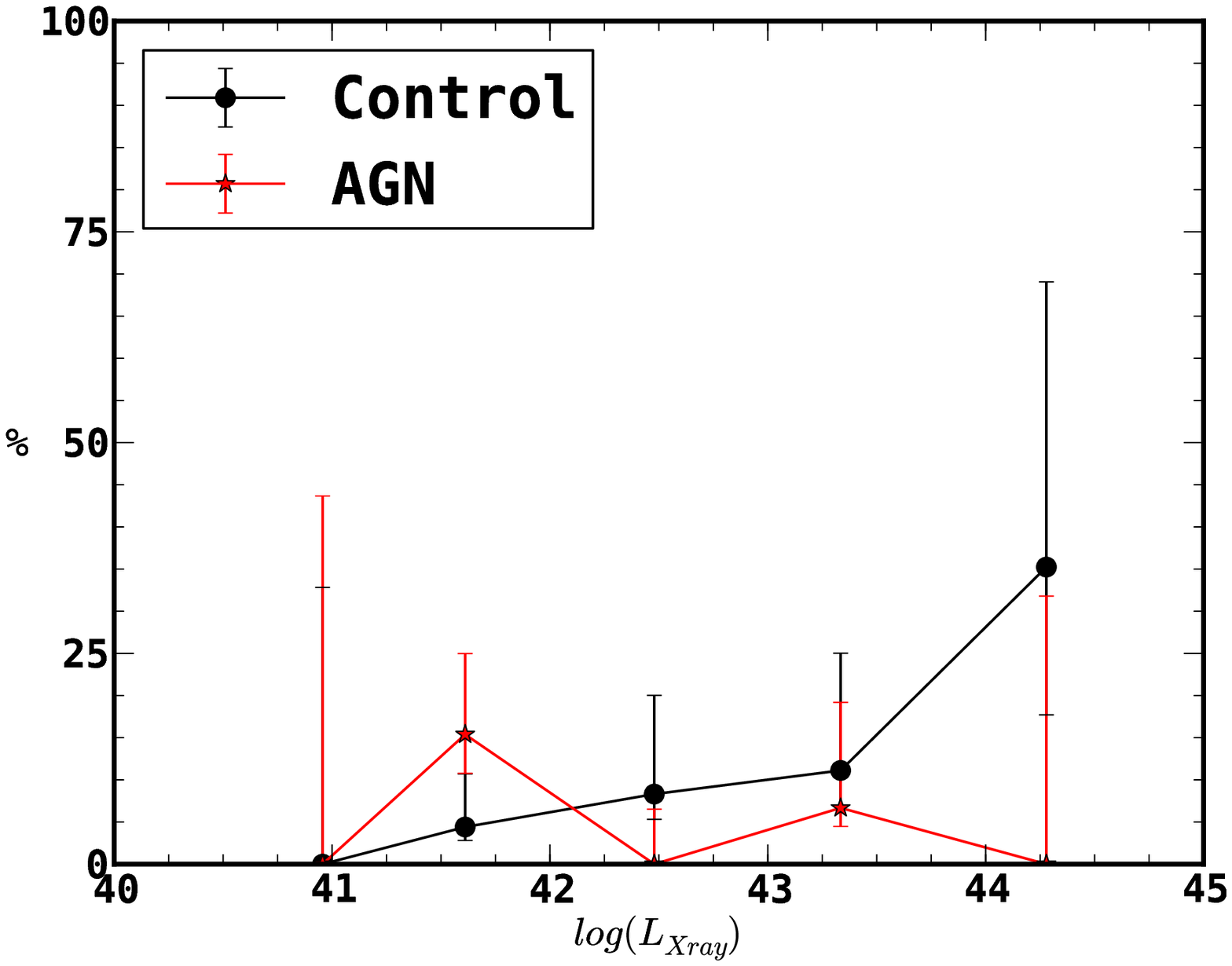}
\caption{Left panel: Asymmetry of AGN Host galaxies compared to matched control samples. The x-axes shows the absorption corrected 2-8kev X-ray luminosities, for control galaxies, this refers to the AGN they were matched to. The y-axes shows the logarithm of the asymmetry A. Histograms of both values are shown in projection on the projected axes. Right panel: Percentage of objects in both AGN sample (red) and control (black) having disturbed morphology (A>0.1). Error bars show one $\sigma$ confidence intervals derived using a beta distribution.}
\label{F:asym}
\end{center}
\end{figure}

The asymmetry measures of AGN host galaxies and matched control galaxies are shown in Fig. \ref{F:asym}, left panel. We find that the level of disturbance in AGN host galaxies is no higher than in the control population. Kolmogorov-Smirnoff tests show statistical differences between the two samples, which appears to be driven by higher skew in the AGN host galaxy sample compared to control. However, the sample sizes studied here are small and larger sample sizes will be needed to analyze this in detail. We also analyze if the incidence of high asymmetry is higher as AGN Xray luminosity rises, to asses this, we set a cut-off limit of $A=0.1$ over which objects are considered disturbed and compare these rates between AGN and matched control, we find no difference between the two samples and no stronger discrepancy at higher Xray luminosities (see Fig. \ref{F:rates}, right panel).
\section{Discussion \& Conclusions}

We have performed a quantitative study of the asymmetries of low redshift (z=0.5-0.8), low luminosity AGN compared to control galaxies. We find no significantly higher asymmetries in the AGN hosts compared to control. This is in good agreement with previous studies that found no strong link between AGN activity and excess signs of merging \cite{cisternas_1,cisternas_2,kocevski,boehm}. This  strengthens the general finding that the majority of low to moderate luminosity AGN are not connected to mergers and the strong merger-AGN link postulated by theoretical models remains undetected (e.g. \cite{hopkins}).

The lacking finding of disturbed morphologies in AGN hosts compared to control can either indicate that merger triggering is limited to the most luminous AGN (as indeed postulated by some theoretical models, e.g. \cite{hopkins_five}). Yet this means that the vast majority of the AGN population shows no connection to mergers. Another possible explanation for our findings and those of others is that there is a long delay between the merger and AGN activity. While such a delay is possible since merger features tend to fade quickly (e.g. \cite{lotz_1,lotz_2}), it causes problem for theoretical models that require AGN to quench star formation soon after a merger has happened \cite{rachel_sam}. An alternate explanation is that X-ray selection detects only relatively unsobscured AGN after the postulated 'blow-out'. We however find no strong trend between asymmetry and X-ray spectral index (a crude estimate for level of obscuration). While some studies find high incidences of obscured AGN activity in young starbursts (e.g. \cite{juneau}), these AGN are often of low luminosity, so it remains unclear how this fits with the general picture of luminous AGN being connected to mergers. 

Further studies of the morphologies of AGN host galaxies, especially at the high luminosity end, are clearly needed. Our understanding of host properties of luminous AGN remains poor, also due to the challenges in studying the hosts of luminous AGN. Theoretical models might have to be adjusted to account for the fact that the vast majority of the AGN population shows no sign of being connected to ongoing mergers and therefore are likely not responsible for quenching starbursts triggered during mergers. These findings also raise the important question of what triggers AGN activity over a wide range of AGN luminosities when mergers are not responsible.

\end{document}